\documentclass{aa}
\usepackage{graphicx}
\usepackage{txfonts}
\usepackage{natbib}
\bibpunct{(}{)}{;}{a}{}{,}

\newcommand{\rosat} {\object{RX J0440.9+4431}}
\newcommand{\ls}    {\object{LS V +44 17}}

\def\simless{\mathbin{\lower 3pt\hbox
     {$\rlap{\raise 5pt\hbox{$\char'074$}}\mathchar"7218$}}}   
\def\simmore{\mathbin{\lower 3pt\hbox
     {$\rlap{\raise 5pt\hbox{$\char'076$}}\mathchar"7218$}}}   

\begin{document}


\title{Long-term optical/IR variability of the Be/X-ray binary \\ \ls/\rosat}

\subtitle{}

\author{P. Reig \inst{1}
\and I. Negueruela\inst{2}
\and J. Fabregat\inst{3}
\and R. Chato\inst{3}
\and M.J. Coe\inst{4}
}

\institute{
IESL (FORTH) \&  University of Crete, Physics Department, PO Box 2208, 710 03
Heraklion, Crete, Greece \\
\email{pau@physics.uoc.gr}
\and Departamento de F\'{\i}sica, Ingenier\'{\i}a de Sistemas y Teor\'{\i}a
de la Se\~nal, Universidad de Alicante, E-03080 Alicante, Spain\\
\and Observatorio Astron\'omico, Universitat de Valencia, 
46071 Paterna-Valencia, Spain\\
\and School of Physics and Astronomy, The University, Southampton, SO17 1BJ,
UK \\
}

\authorrunning{Reig et~al.}
\titlerunning{Optical/IR observations of \rosat}

\offprints{P. Reig, \\ \email{pau@physics.uoc.gr}}

\date{Received / Accepted }

\abstract{

We present the first long-term study of the optical counterpart to the
X-ray pulsar \rosat/\ls. The data consist of optical spectroscopic and
infrared photometric observations taken during the period 1995-2005. The
infrared observations are the first published for this source. The results
of our photometric and spectroscopic analysis show that  \rosat/\ls\
contains a moderately reddened, E(B-V)=0.65$\pm$0.05, B0.2V star located
at about 3.3 kpc.  The H$\alpha$ line consistently 
shows a double-peak profile
varying from symmetric shape to completely distorted on one side
(V/R phases). A correlation between the equivalent width of the H$\alpha$
line and the infrared magnitudes is seen: as the  EW(H$\alpha$) decreases
the IR magnitudes become fainter. This long-term optical/IR variability is
attributed to structural changes in the Be star's circumstellar disc. The
observations include a recent decline in the
circumstellar disc and subsequent recovery.   We have witnessed the
cessation of a global oscillation due to the decline of the circumstellar
disc. If the present disc growth rate continues we predict the onset of
another episode of V/R variability by the end of 2006. We have
investigated the typical time scales for disc variability of various
Be/X-ray binaries and found a correlation with the orbital period. This
correlation is hard to establish due to the difficulty in defining the exact
duration of the various activity states, but it is seen both in the
duration of the disc growth/dissipation phase and the value of the
H$\alpha$ equivalent width prior to the appearance of asymmetric
profiles. These relationships provide further evidence for the interaction
of the neutron star with the circumstellar disc of the Be star's companion
and confirms the need of a fully developed disc for the V/R variability to
be observed.

\keywords{stars: individual: \object{RX J0440.9+4431}, \object{LS V +44 17}
 -- X-rays: binaries -- stars: neutron -- stars: binaries close --stars: 
 emission line, Be}
}

\maketitle

\begin{figure*}
\begin{center}
\includegraphics[width=16cm,height=7cm]{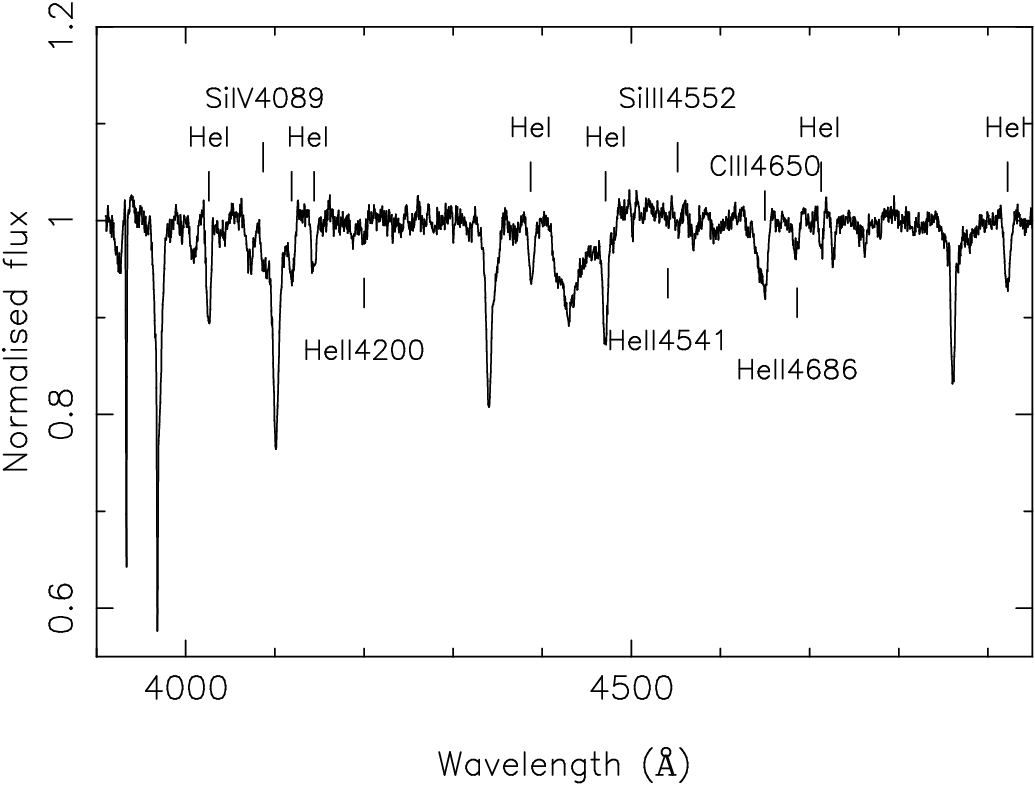}
\caption[]{Blue spectrum showing the lines used for spectral
classification. The spectrum was taken from the INT on October 25, 2002. The
\ion{He}{i} lines marked are $\lambda$4026, $\lambda$4121, $\lambda$4144, 
$\lambda$4387, $\lambda$4471, $\lambda$4713 and $\lambda$4921}
\label{sp}
\end{center}
\end{figure*}

\section{Introduction} \label{introduction}

\ls\ is a relatively bright V=10.8 B0 star that is associated with the
X-ray source \rosat. \rosat/\ls\ belongs to the subgroup of high-mass
X-ray binaries known as Be/X-ray binaries. These systems consist of a
neutron star orbiting a O9e-B2e star. The letter $e$ stands for emission,
as instead of the normal photospheric absorption lines the optical spectra
of Be stars display emission lines. Although helium and iron are
occasionally seen in emission, the hydrogen lines, especially those of the
Balmer series, constitute the signature for which Be stars are renowned
\citep{port03}.
Strong infrared emission is another defining characteristic of Be stars. 
The origin of these two observational properties (emission lines and
infrared excess) resides in a gaseous, equatorially concentrated
circumstellar disc around the OB star. This disc acts as a reservoir of
material for accretion on to the compact object. Although the
ultimate cause of the Be phenomenon is still not known, it is believed to
be related to the rapid rotation of these stars. Recent studies
\citep{town04} indicate that Be stars may rotate much closer to break-up
velocity that previously thought.

The optical/IR information on \ls\ is very scarce. It mainly comes from
surveys of the Galactic Plane or catalogues prepared for specific space
missions. However, no variability studies of this system have ever been
reported. Interest in \ls\ grew when the first evidence
that it might be an X-ray binary came to light \citep{motc97}. 
\rosat\ is an X-ray pulsar with a spin period of 202 seconds and belongs
to the poorly studied group of persistent Be/X-ray binaries \citep{reig99}.

In this work we present the results of our monitoring programme on
high-mass X-ray binaries for \rosat/\ls. We have performed a detailed
analysis of its optical and infrared variability covering a period of
almost 10 years. We have also made a comparative study of the
long-term variability time scales of various Be/X-ray binaries.

\begin{table*}
\begin{center}
\caption{Log of the spectroscopic observations.}
\label{log}
\begin{tabular}{cccccccc}
\hline \hline \noalign{\smallskip}
\multicolumn{8}{c}{Red-end spectra} \\
\noalign{\smallskip} \hline \noalign{\smallskip}
Date	&Telescope/	&Dispersion	&Grating  &Wavelength	&MJD	&EW(H$\alpha$)	&$\log(\rm{V/R})^a$\\
	&Observatory	&\AA/pixel	&l/mm	 &\AA		&	&\AA		&\\
\noalign{\smallskip} \hline \noalign{\smallskip}
29-11-1995	&1.0m/ORM	&1.07	&1200	&6100-6960	&50050.5	&--9.3$\pm$0.4	&0.07	\\
28-02-1996	&1.0m/ORM	&0.34	&2400	&6400-6750	&50141.5	&--8.8$\pm$0.4	&--0.25\\
26-10-1997	&1.0m/ORM	&0.90	&1200	&5900-6800	&50747.6	&--8.3$\pm$0.3	&0.00\\
28-10-1997	&1.0m/ORM	&0.34	&2400	&6380-6720	&50749.6	&--8.3$\pm$0.4	&--0.02\\
20-07-2000	&1.3m/SKI	&1.04	&1302	&5550-7550	&51745.6	&--4.4$\pm$0.2	&--0.03\\
05-10-2000	&1.3m/SKI	&1.04	&1302	&5250-7290	&51822.6	&--4.4$\pm$0.1	&0.04\\
07-08-2001	&1.3m/SKI	&1.04	&1302	&5660-7440	&52128.6	&--0.2$\pm$0.1	&0.06\\
12-09-2001	&1.3m/SKI	&1.04	&1302	&5220-7190	&52164.6	&--0.5$\pm$0.1	&--0.03\\
13-09-2001	&1.3m/SKI	&1.04	&1302	&5230-7200	&52165.6	&--0.5$\pm$0.1	&--0.02\\
08-10-2001	&1.3m/SKI	&1.04	&1302	&5460-7430	&52190.6	&--1.1$\pm$0.1	&--0.04\\
22-10-2001	&1.9m/OHP	&0.44	&1200	&6250-7150	&52205.6	&--1.0$\pm$0.1	&0.03\\
05-12-2001	&2.5m/NOT	&1.5	&600	&3800-6800	&52248.5	&--1.4$\pm$0.1	&--0.02\\
20-01-2002	&1.5m/OHP	&0.22	&600	&6380-6830	&52295.2	&--1.9$\pm$0.1	&--0.04\\
25-07-2002	&2.5m/ORM	&0.8	&1200	&6270-7120	&52480.7	&--1.7$\pm$0.1	&--0.04\\
10-09-2002	&1.3m/SKI	&1.04	&1302	&4680-6750	&52527.6	&--1.5$\pm$0.1	&0.01\\
28-10-2002	&2.5m/ORM	&1.4	&400	&3400-7400	&52575.7	&--2.0$\pm$0.1	&--0.04\\
06-10-2003	&1.3m/SKI	&1.04	&1302	&5260-7330	&52918.6	&--6.1$\pm$0.2	&--0.03\\
02-02-2004	&4.2m/ORM	&0.22	&1200	&6100-6800	&53038.4	&--6.0$\pm$0.2	&0.01\\
25-08-2004	&1.3m/SKI	&1.04	&1302	&4770-6840	&53243.6	&--6.1$\pm$0.2	&0.03\\
27-08-2004	&1.3m/SKI	&1.04	&1302	&4770-6840	&53245.5	&--6.0$\pm$0.2	&--0.03\\
24-10-2004	&1.3m/SKI	&1.04	&1302	&4770-6840	&53303.4	&--5.9$\pm$0.2	&0.00\\
25-10-2004	&1.3m/SKI	&1.04	&1302	&4770-6840	&53304.6	&--5.9$\pm$0.2	&--0.03\\
19-02-2005	&4.0m/KPNO	&1.74	&316	&4940-9290	&53420.7	&--5.8$\pm$0.2	&--0.01\\
\hline \hline \noalign{\smallskip}
\multicolumn{8}{c}{Blue-end spectra} \\
\noalign{\smallskip} \hline \noalign{\smallskip}
Date	&Telescope/	&Dispersion	 &Grating &Wavelength	&MJD	&EW(H$\beta$)	&profile\\
	&Observatory	&\AA/pixel	 &l/mm   &\AA		&	&\AA		& \\
\noalign{\smallskip} \hline \noalign{\smallskip}
01-03-1996	&1.0m/ORM	&0.44	&1200	&4550-5000	&50143.5	&--0.5		&shell	\\
24-10-1997	&1.0m/ORM	&0.90	&1200	&4000-4950	&50745.7	&--0.4		&shell	\\
22-07-2000	&1.3m/SKI	&1.04	&1302	&3800-5700	&51747.6	&+0.6		&absorption	\\
17-10-2000	&1.3m/SKI	&1.04	&1302	&3600-5600	&51834.6	&+0.7		&absorption	\\
09-08-2001	&1.3m/SKI	&1.04	&1302	&3900-5900	&52130.6	&+1.6		&absorption	\\
18-09-2001	&1.5m/OHP	&0.22	&600	&4440-4900	&52170.7	&+1.7		&absorption	\\
07-10-2001	&1.3m/SKI	&1.04	&1302	&3900-5900	&52189.6	&+1.6		&absorption	\\
23-10-2001	&1.9m/OHP	&0.9	&600	&3745-5575	&52205.5	&+2.0		&absorption	\\
25-10-2002	&2.5m/ORM	&0.5	&1200	&3600-5400	&52572.7	&+1.4		&absorption	\\
02-02-2004	&4.2m/ORM	&0.22	&1200	&4000-4800	&53038.4	&--		&--	\\
\noalign{\smallskip} \hline \hline
\multicolumn{8}{l}{$a$: $V/R=(I(V)-I_c)/(I(R)-I_c)$}
\end{tabular}
\end{center}
\end{table*}
\begin{table*}
\begin{center}
\caption{Optical and infrared magnitudes of \ls.}
\label{phot}
\begin{tabular}{lcccccc}
\hline \hline \noalign{\smallskip}
\multicolumn{7}{c}{Optical} \\
\noalign{\smallskip} \hline \noalign{\smallskip}
Date	&MJD	&Observatory	&$y$		&$b$		&$v$	&$u$  \\
\noalign{\smallskip} \hline \noalign{\smallskip}
16-08-1999 &51407.50 &1.3m/SKI &10.82$\pm$0.03 &11.38$\pm$0.03 &11.64$\pm$0.05 &12.13$\pm$0.06\\
\noalign{\smallskip} \hline \noalign{\smallskip}
	&	&	&$B$		&$V$		&$R$	&$I$  \\
\noalign{\smallskip} \hline \noalign{\smallskip}
24-08-2004 &53242.59 &1.3m/SKI &11.48$\pm$0.02 &10.85$\pm$0.02 &10.40$\pm$0.02 &9.92$\pm$0.02 \\
14-09-2004 &53263.56 &1.3m/SKI &11.39$\pm$0.02 &10.75$\pm$0.02 &10.30$\pm$0.02 &9.83$\pm$0.02 \\
01-10-2004 &53280.51 &1.3m/SKI &11.40$\pm$0.02 &10.76$\pm$0.02 &10.33$\pm$0.02 &9.85$\pm$0.02 \\
\noalign{\smallskip} \hline \hline
\multicolumn{7}{c}{Infrared} \\
\noalign{\smallskip} \hline \noalign{\smallskip}
Date	&MJD	&Instrument$^a$	&$J$		&$H$		&$K$	&\\
\noalign{\smallskip} \hline \noalign{\smallskip}
14-10-95 & 50005.71 &CVF & 8.99$\pm$0.01 & 8.73$\pm$0.01 & 8.44$\pm$0.01\\
12-01-96 & 50095.58 &CVF & 9.06$\pm$0.01 & 8.76$\pm$0.01 & 8.49$\pm$0.01\\
27-10-98 & 51114.62 &CVF & 9.53$\pm$0.01 & 9.47$\pm$0.04 & 9.21$\pm$0.01\\
27-10-98 & 51114.63 &CVF & 9.60$\pm$0.01 & 9.40$\pm$0.02 & 9.26$\pm$0.01\\
28-10-98 & 51115.61 &CVF & 9.40$\pm$0.01 & 9.18$\pm$0.01 & 8.99$\pm$0.01\\
02-10-99 & 51454.59 &CVF & 9.40$\pm$0.02 & 9.21$\pm$0.02 & 8.99$\pm$0.02\\
22-01-00 & 51566.48 &CVF & 9.51$\pm$0.03 & 9.36$\pm$0.02 & 9.18$\pm$0.02\\
22-01-00 & 51566.49 &CVF & 9.56$\pm$0.02 & 9.30$\pm$0.01 & 9.23$\pm$0.02\\
17-10-00 & 51835.73 &CVF & 9.57$\pm$0.03 & 9.45$\pm$0.02 & 9.34$\pm$0.03\\
11-01-01 & 51921.51 &CVF & 9.57$\pm$0.03 & 9.46$\pm$0.02 & 9.37$\pm$0.02\\
14-01-01 & 51924.48 &CVF & 9.56$\pm$0.02 & 9.40$\pm$0.02 & 9.33$\pm$0.03\\
23-03-02 & 52357.40 &CAIN-II & 9.47$\pm$0.03 & 9.18$\pm$0.03 & 9.07$\pm$0.03\\
08-11-02 & 52587.78 &CAIN-II & 9.10$\pm$0.03 & 8.89$\pm$0.03 & 8.73$\pm$0.03\\
09-11-02 & 52588.67 &CAIN-II & 9.15$\pm$0.03 & 8.92$\pm$0.03 & 8.77$\pm$0.03\\
12-11-02 & 52591.54 &CAIN-II & 9.26$\pm$0.03 & 8.93$\pm$0.03 & 8.75$\pm$0.03\\
13-11-02 & 52592.67 &CAIN-II & 9.28$\pm$0.03 & 8.95$\pm$0.03 & 8.79$\pm$0.03\\
30-08-04 & 53248.69 &CAIN-II & 9.24$\pm$0.03 & 8.83$\pm$0.03 & 8.71$\pm$0.03\\
29-12-04 & 53369.57 &FIN & 9.25$\pm$0.10 & 9.03$\pm$0.03 & 8.72$\pm$0.04\\
30-12-04 & 53370.57 &FIN & 9.19$\pm$0.04 & 9.03$\pm$0.03 & 8.73$\pm$0.04\\
\noalign{\smallskip} \hline \hline
\multicolumn{7}{l}{$a$: All observations were obtained from the 1.5-m Carlos
S\'anchez Telescope}
\end{tabular}
\end{center}
\end{table*}

\section{Observations}

\subsection{Optical observations}

Optical spectroscopic observations were obtained from 7 telescopes at 4
different observatories: from the Roque de los Muchachos observatory (ORM)
in La Palma (Spain), observations were made with the 1.0m Jacobus Kapteyn
Telescope, the 2.5m Isaac Newton Telescope, the 4.2m William Herschel
Telescope (service time) and the 2.5m Nordic Optical Telescope; from the
Skinakas observatory (SKI) in Crete (Greece) the data come from the 1.3m
telescope; and from the Haut Provence observatory (OHP) in France the 1.52m
and the 1.93m telescopes were employed. Finally one spectrum was taken
from the 4m telescope of the Kitt Peak National Observatory (KPNO) in the
USA. Table~\ref{log} gives the log of the spectroscopic observations. This
table contains instrumental information together with the results of the
spectral analysis: the equivalent width of the H$\alpha$ and H$\beta$
lines and an indication of the profile shape of the lines. Negative values
indicate that the line is in emission.  The reduction of the spectra was
made using the STARLINK {\em Figaro} package \citep{shor01}, while their
analysis was performed using the STARLINK {\em Dipso} package
\citep{howa98}.

Optical photometric observations were made using two photometric systems.
Str\"omgren photometry ($uvby$) was obtained from the Skinakas observatory
(SKI) on August 16, 1999. \ls\ was also
observed through the  Johnson $B$, $V$, $R$ and $I$ filters on three
occasions from the Skinakas observatory (see Table~\ref{phot}). The
telescope was equipped with a 1024$\times$1024 SITe CCD chip, containing
24$\mu$m pixels. Reduction of the data was carried out using the IRAF
tools for aperture photometry.

\subsection{Infrared observations}

Infrared photometry in the JHK bands was obtained as part of a
monitoring programme of Be/X-ray binaries at the 1.5 m. Carlos S\'anchez
Telescope (TCS), located at the Teide Observatory in Tenerife, Spain.
The instruments used were the Continuously Variable Filter Photometer
(CVF) up to January 2001, and the CAIN-II camera, equiped with a 
$256\times 256$ HgCdTe (NICMOS 3) detector ever since. The last data in
December 2004 were obtained with the recently comissioned FIN photometer.

Instrumental CVF and FIN magnitudes were transformed to the standard
system defined by \citet{kidg03}. Instrumental CAIN magnitudes were
obtained from the images by means of the IRAF tools for aperture
photometry, and transformed to the standard system defined by
\citet{hunt98}. The accuracy of the standard JHK values in all three bands
is  0.01, 0.03 and 0.02 mag. for CVF, CAIN-II and FIN data respectively. 
The obtained values are given in Table~\ref{phot}.

\section{Results}

\subsection{Previous optical work}

The first astronomical observations date back to the circa 1930. 
\ls\ is
mentioned in the {\em Bergedorfer Spektral-Durchmusterung} catalogue 
\citep[BSD,][]{schw35}, where a spectral type B0 is suggested. The first
accurate photometric observations on the Johnson photometric system were
performed by \citet{biga63} who gave $V=10.78$, $(B-V)=0.61$ and
$(U-B)=-0.36$ for observations performed in 1953, although the star is
referenced in \citet{seyf41} with a photographic magnitude of 11.3. \ls\ is
also found on one of the photographic plates of the Sandage two-colour
survey of the Galactic plane \citep{lann01} with $B=10.4$ and
$(U-B)=-0.4$, although these values are affected by large errors
($\simmore \pm 0.5$ mag) owing to the uncertainties inherent to obtaining
accurate visual estimates from photographic plates. 

Further photometric data are provided by catalogues of optical surveys.
However, varying results are obtained for different versions of
the catalogue due to slightly different reduction methods. Another
disadvantage of the catalogued values is the difficulty in defining the
exact date of the observations.  The Hipparcos and Tycho Catalogues
\citep{hog00} give $V=10.71$ and $(B-V)=0.44$ for observations that took
place some time from November 1989 to March 1993, whereas the USNO-A2.0
Catalogue \citep{mone98} gives a blue magnitude of 12.2 and a red
magnitude of 10.4 for the epoch 1953.025.

\subsection{Spectral classification}

The presence of \ion{H}, \ion{He}{i} and \ion{Si}{iv} lines clearly
demonstrates that \rosat\ contains an early-type star (Fig.~\ref{sp}).
He\,II lines, although present, are weak (only \ion{He}{ii}~4686\AA\ is
clearly detected, \ion{He}{ii}~4200\AA\ is weak and \ion{He}{ii}~4541\AA\
is buried in the noise), which implies that the spectral type must be
earlier than B1 but later than O9.5. A spectral type in this range is also
borne out by the weak \ion{Si}{iii}~4552\AA.  \ion{He}{ii}~4686\AA\ is
normally last seen at B0.5, whereas the ratio
\ion{Si}{iii}~4552\AA/\ion{Si}{iv}~4089\AA\ increases smoothly as we
progress toward later spectral types. A visual comparison of the relative
strength of metallic lines of \ls\ with those of MK standards
\citep{walb90} resulted in B0.2 as the closest spectral type. In
particular, the fact that the ratio of the \ion{C}{iii}~4650\AA\ blend
with respect to \ion{He}{ii}~4686\AA\ is larger than one agrees more with a
B0.2 than with a B0 type. As for the luminosity class, a main-sequence
star is suggested by the fact that the ratios of \ion{He}{i}~4026\AA\ over
\ion{Si}{iv}~4089\AA\ and \ion{He}{i}~4121\AA\ over \ion{Si}{iv}~4116\AA\
are larger than one. Therefore, we conclude that the optical counterpart to
the Be/X-ray binary \rosat\ is a B0.2V star. 

\begin{figure}
\resizebox{\hsize}{!}{\includegraphics{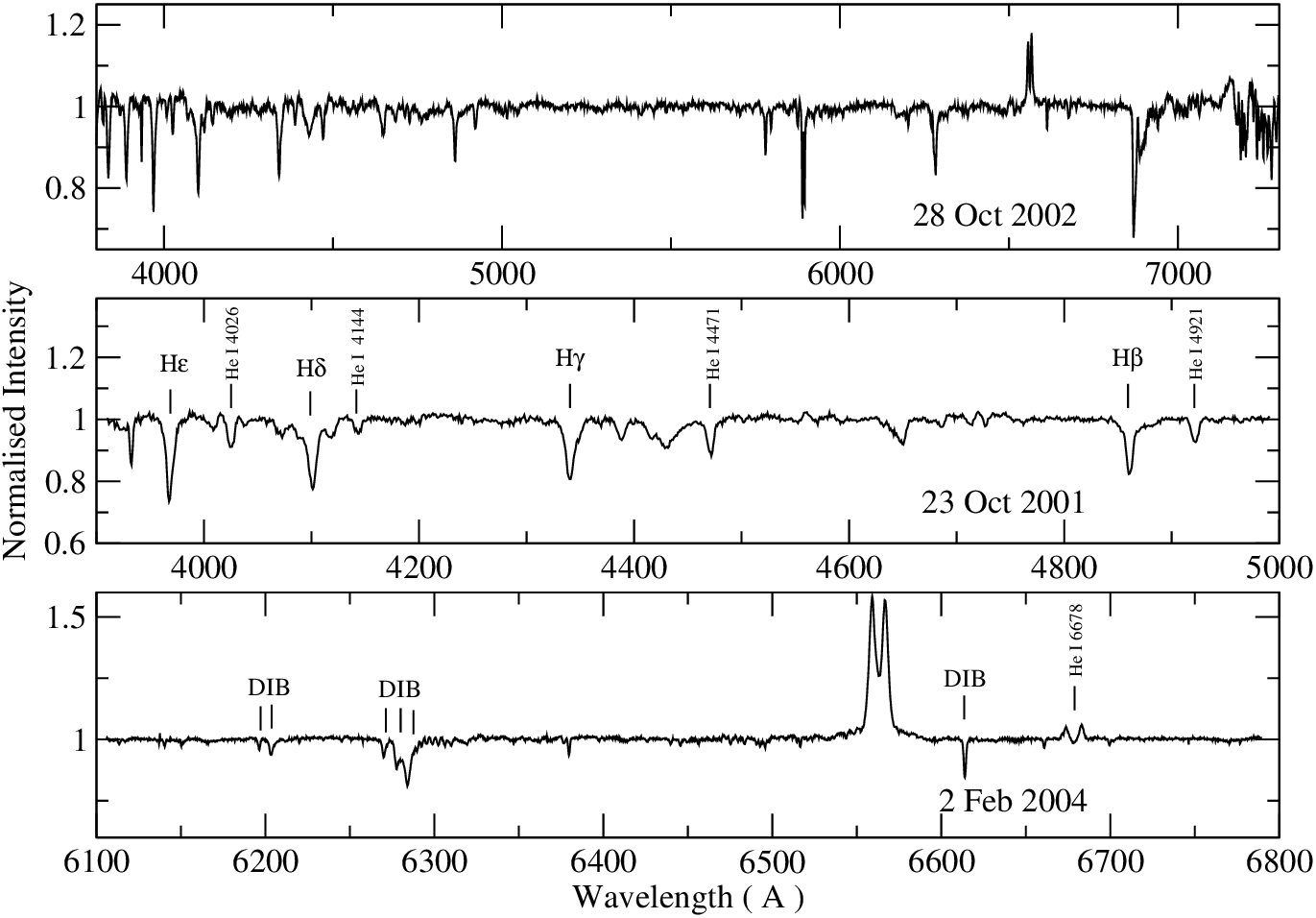}}
\caption[]{{\it top:} Broad-band low-resolution spectrum of the optical 
companion of \rosat. {\it middle {\rm and} bottom:} Blue-end and red-end spectra 
obtained with higher resolution. The most relevant hydrogen and helium lines
and diffuse interstellar bands are marked. The Balmer series can be seen as 
well.}
\label{redblue}
\end{figure}
\begin{figure*}
\resizebox{\hsize}{!}{\includegraphics{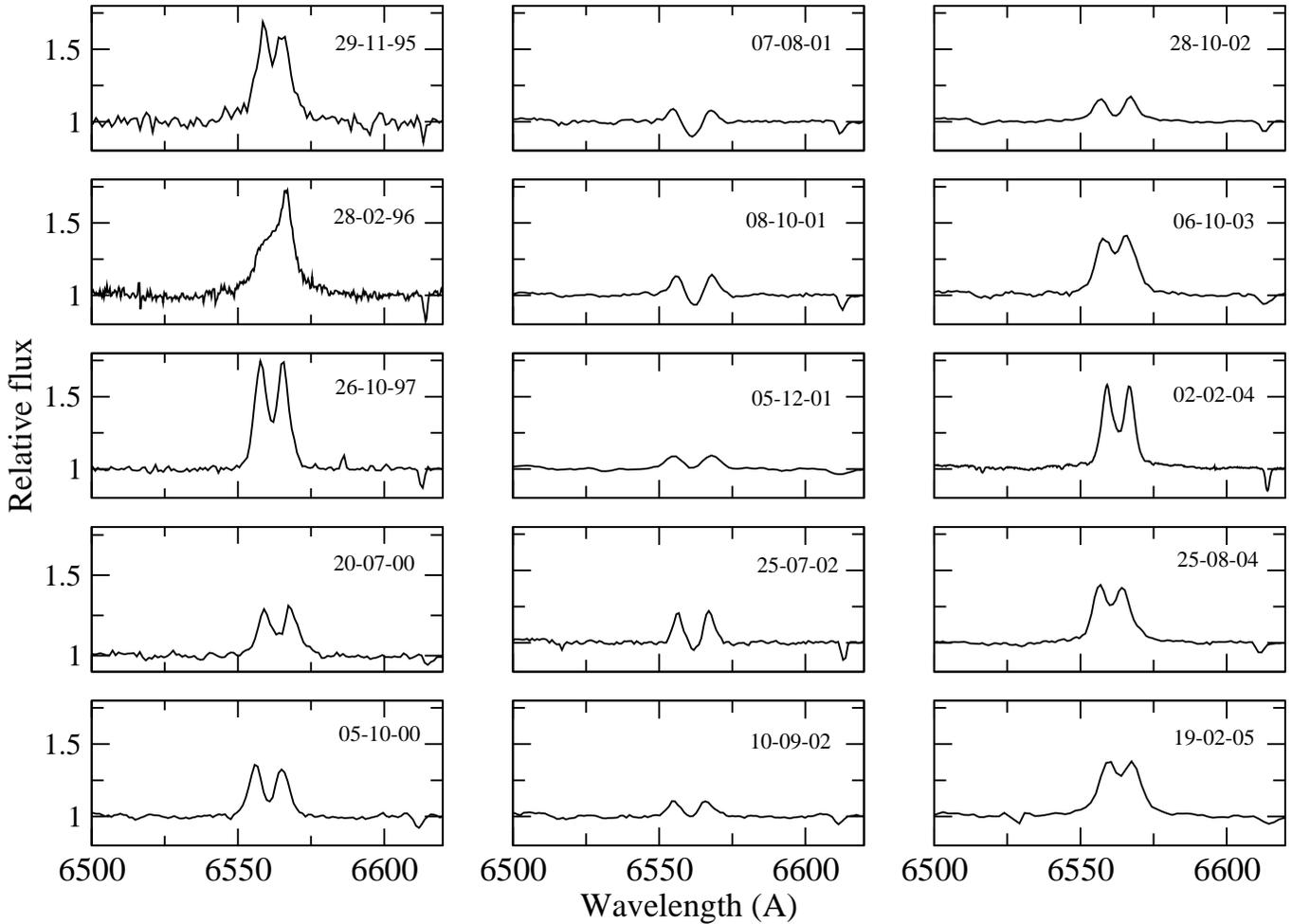}}
\caption[]{Evolution of the H$\alpha$ line profile for the past 10 years.
See Table~\ref{log} for the complete set of observations.}
\label{havar}
\end{figure*}

\subsection{Reddening and distance}

Besides the H$\alpha$ and He\,I $\lambda6678$ lines, the red-end spectrum
(Fig.\ref{redblue}) contains several strong diffuse interstellar bands
(DIB), which can be used to estimate the amount of interstellar absorption
toward the source \citep{herb91,gala00}.  We used six
different interstellar lines (6010 \AA, 6195 \AA, 6202 \AA, 6269 \AA,
6376/79 \AA\ and 6613 \AA) to derive a colour excess $E(B-V)$ according to
the linear relationships of \citet{herb75}. Given the different amount
of available data (the spectra covered different wavelength intervals) and
different signal-to-noise values of the spectra, we obtained a mean
$E(B-V)$ by averaging all measurements. The resulting reddening was
$E(B-V)=0.62\pm0.03$. For the sake of comparison, we also obtained the
colour excess using only the average of the five measurements of the
highest resolution spectra (those taken from the WHT on February 2, 2004).
The resulting $E(B-V)$ was 0.64$\pm$0.02. The errors are the weighted
standard deviation of the results of the various lines used.

Estimating the reddening from photometric data in Be stars might be
misleading as the circumstellar continuum emission affects the photometric
colours and indices \citep{fabr98}. This effect is
expected to be more distinct for longer wavelengths. Indeed,  a B0.2V star
has an intrinsic colour $(B-V)_0=-0.25$ \citep{wegn94}. Taking the
measured photometric magnitudes $(B-V)=0.63$ (2004 observations) we derive
an excess $E(B-V)\approx 0.9$, somewhat larger than the value derived
above. However, \ls\ went through a low-activity optical state, presumably
during the first half of 2001 as indicated by the low H$\alpha$ equivalent
widths (see Fig.~\ref{ewir}). Interpreting this optical minimum as a weakening of the disc we
would expect that the IR magnitudes at that time should be very close to
those of the underlying B star. Assuming the interstellar extinction law
$E(J-K)=0.54E(B-V)$ and the intrinsic colour $(J-K)_0=-0.16$ for a
main-sequence B0.2 star \citep{koor83}, the observed
$(J-K)=0.21$ gives $E(B-V)=0.68\pm0.03$, in good agreement with the
spectroscopically derived value. 

Finally, taking the standard law $A_V=3.1 E(B-V)$ and assuming an average
absolute magnitude for a B0.2V star of $M_V=-3.8$ \citep{hump84,mart05} the
distance to \rosat\ is estimated to be $\sim$ 3.3$\pm$0.5 kpc.  This error
includes those of $m_V$ (0.02) and $A_V$ (0.3), but assumes no error
in the absolute magnitude $M_V$.

\subsection{Rotational velocity}

The rotational velocity was estimated by measuring the full width at half
maximum of He I lines \citep[see e.g.][]{stee99}. After correcting for
instrumental resolution we obtained $v \sin i=235\pm15$ km s$^{-1}$, which
compares favourably 
to the value of 246$\pm$16 km s$^{-1}$ given for weak-emission
early-type shell stars \citep{menn94}. As a comparison, other rotational
velocities in Be/X-ray binaries are:  $v \sin i=200\pm30$ km s$^{-1}$ in
\object{LS I +61 235/RX J0146.9+6121} \citep{reig97a}, $v \sin i=290\pm50$
km s$^{-1}$ in \object{V635 Cas/4U 0115+63} \citep{negu01}, $v \sin
i=240\pm20$ km s$^{-1}$ in  \object{LS 992/RX J0812.4--3114}
\citep{reig01}, $v \sin i=240\pm20$ km s$^{-1}$ in \object{SAX
J2103.5+4545} \citep{reig04}.

\begin{figure}
\includegraphics[width=8cm,height=7cm]{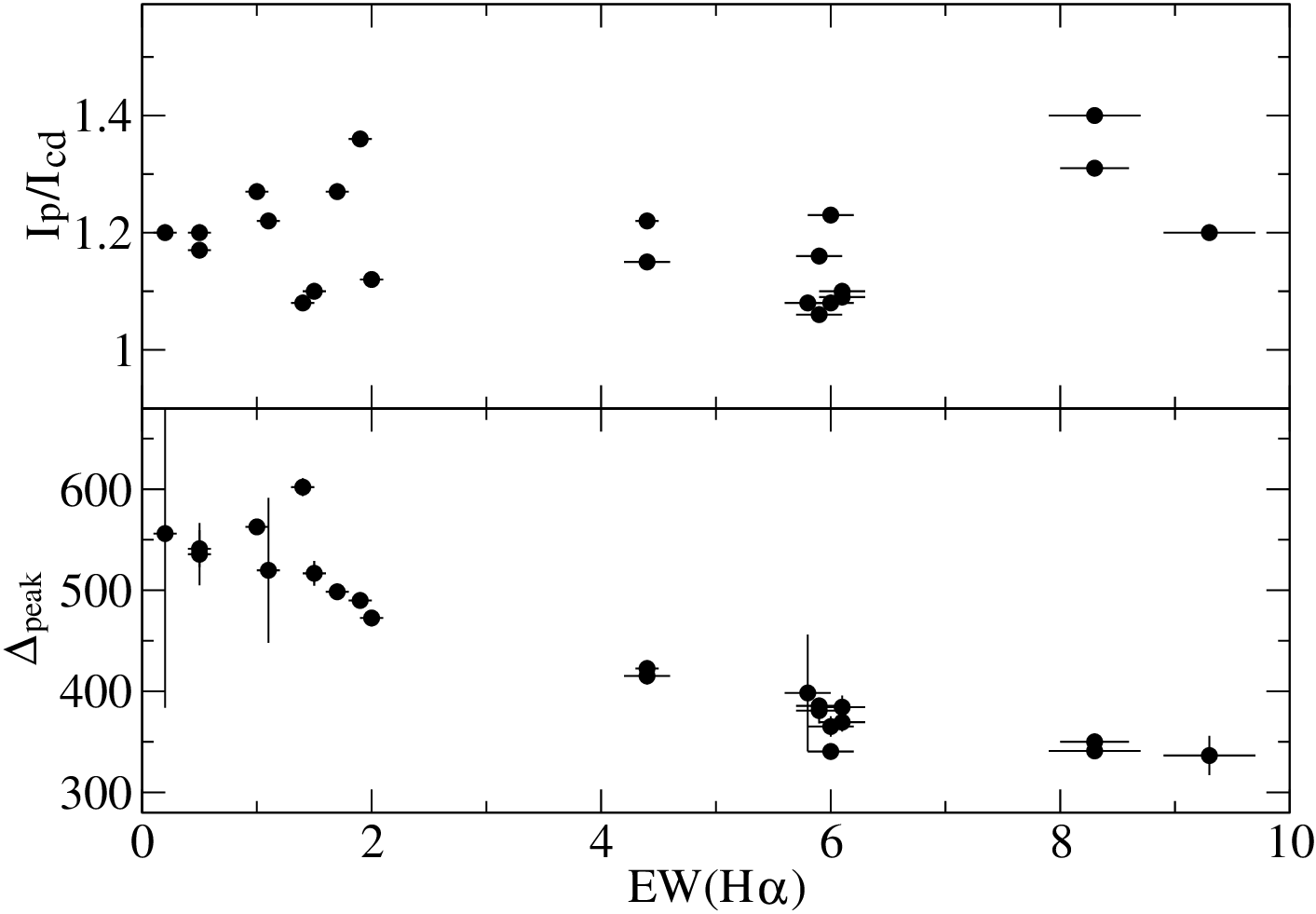}
\caption[]{{\it Top}: mean intensity of the blue and red peaks $I_{\rm p}$ 
over that of the central reversal $I_{\rm cd}$. {\it Bottom}: peak 
separation as a function of the H$\alpha$ equivalent width. Note that the
spectral resolution constitutes a source of scattering in these diagrams.}
\label{havarew}
\end{figure}

\section{Discussion}

\begin{figure}
\includegraphics[width=8cm,height=7cm]{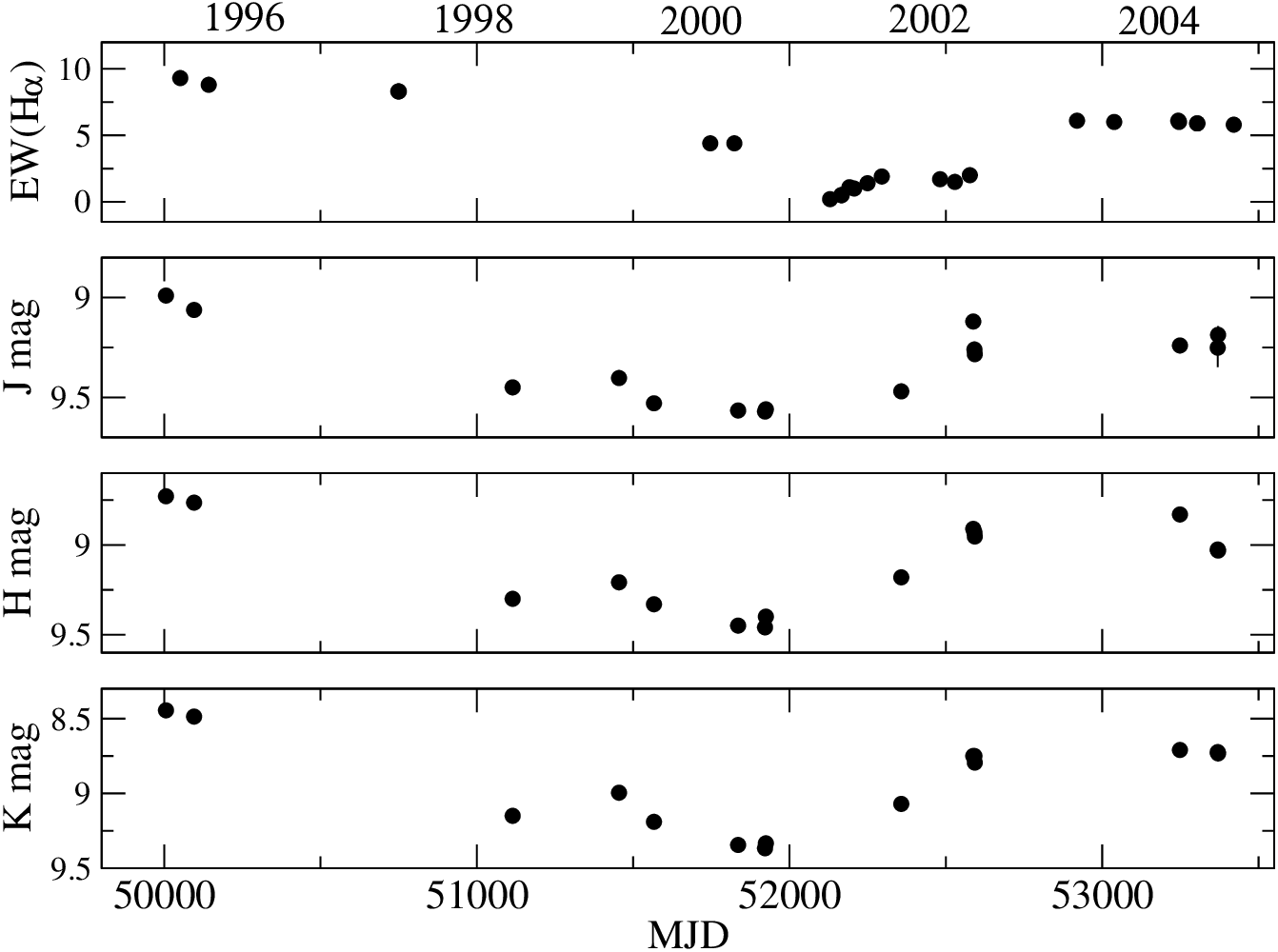}
\caption[]{Evolution of the H$\alpha$ equivalent width and IR magnitudes 
for the past 10 years. Errors are included in the size of the points.}
\label{ewir}
\end{figure}

We have monitored the Be/X-ray binary \ls\ for the last 10 years. Our
observations coincided with the latest stages of a declining disc phase.
The slow and gradual decline of the  EW(H$\alpha$) and IR colours seems to
indicate that the mechanism that feeds the disc had already stopped when
we started the monitoring of the source. The  source entered a long period
(1998-2003) of low optical/IR activity, where the line emission just
filled in the underlying absorption expected from the photosphere of the
B-type star and the IR magnitudes showed their lowest values.  The
equivalent width of the H$\alpha$ line (EW(H$\alpha$)) always remained
negative, indicating that the complete loss of the disc did not
occur. However, given the large observational gaps of our data we
cannot rule out the possibility 
that such an event could have happened. The loss of the
circumstellar disc could have occurred in early 2001 (the EW(H$\alpha$) was
only --0.2 \AA\ in August 2001). As mentioned before, in January 2001 the
measured intrinsic IR colour $(J-K)\approx -0.16$  agrees with a
B0-B0.5 star \citep{koor83}. In other words, in 2001 the underlying B-type
star would have been exposed. Figure \ref{ewir} shows the evolution of the
H$\alpha$ equivalent width and the infrared magnitudes.  The 
EW(H$\alpha$) and the IR magnitudes follow the same trend, namely, a slow
decrease, reaching a minimum around MJD 52000 and a gradual increase. This
long-term variability suggests the dissipation and subsequent formation of
the circumstellar disc and sets a common origin (i.e., the circumstellar
disc) for the H$\alpha$ emission and infrared excess.

\begin{table*}
\begin{center}
\caption{Typical time scales for disc variability in Be/X-ray binaries.}
\label{loss}
\begin{tabular}{llccccl}
\noalign{\smallskip} \hline \hline
X-ray	&Optical&Spectral &$T_{\rm disc}$ &$T_{\rm V/R}$ &$P_{\rm orb}$ &Reference\\
name	&name	&type	&(year)		&(year)		&(day)		&\\
\noalign{\smallskip} \hline \hline
4U 0115+63	&V635 Cas	&B0.2Ve		& 3-5	&0.5-1.5	&24.3	 &\citet{negu01}\\
V 0332+53	&BQ Cam		&O9Ve		& 4-5	&1		&34.2	 &\citet{negu99,gora01}\\
4U 0352+309	&X Per		&B0Ve		& 7	&0.6-2		&250	 &\citet{clar01}\\
1A 0535+262	&V725 Tau	&O9.7IIIe	& 4-5	&1-1.5		&111	 &\citet{haig04,clar98}\\
RX J0812.4-3114	&LS 992		&B0.2IVe	& 4	&$-$		&80	 &\citet{reig01}\\		 
RX J 0146.9+6121&LS I +61 235	&B1Ve		&$>$10	&3.4		&$>200^*$  &\citet{reig00}\\
4U 1145-619	&V801 Cen	&B1Ve		&$>$10	&3		&186	 &\citet{stev97}\\
RX J0440.9+4431	&LS V +44 17	&B0.2Ve		&$>$10	&$-$		&$>150^*$  &this work\\
\hline
\multicolumn{7}{l}{$*$ obtained from the $P_{\rm spin}-P_{\rm orb}$
correlation \citep{corb86}} \\
\end{tabular}
\end{center}
\end{table*}

\subsection{The H$\alpha$ line}

V/R variability is defined as the intensity variations of the two peaks
(known as violet and red peak) in the split profile of a spectral line. In
many Be stars, if monitored over a long enough periods of time, these
variations are quasiperiodic \citep{okaz97}. The V/R ratio is defined as
$V/R=(I(V)-I_c)/(I(R)-I_c)$, where $I(V)$, $I(R)$ and $I_c$ are the
intensities of the violet peak, red peak and continuum, respectively.

The evolution of the H$\alpha$ profile throughout the period covered by our
observations is presented in Fig.\ref{havar}. The vertical scale was left
the same in all plots in order to show the variability in the strength of
the line. Double-peak H$\alpha$-line profiles, both symmetric and
asymmetric, are always present in \ls. Symmetric profiles are believed to
be generated in quasi-Keplerian discs \citep[see e.g][]{humm94}. Asymmetric
profiles are associated with radial motion and/or distorted density
distributions \citep{hanu95,humm97}. The model that most successfully
accounts for the long-term variability of these asymmetric profiles is the
one-armed oscillation model  \citep[][and references therein]{okaz00}.

In \ls,  the ratio of the intensities of the violet over the red peak
($V/R$) hardly varied over almost 10 years. Excluding the first two
spectra, the lines have $|\log(V/R)| < 0.05$ (column 8 of
Table~\ref{log}). A symmetric profile does not necessarily mean the
absence of the density wave as symmetric split profiles (the V=R phase)
can occur during a fraction of the V/R cycle, more precisely, when the
star lies between the observer and the high-density perturbation
\citep{telt94}. These V=R phases represent a fraction of the entire
V/R cycle. In \ls\ an asymmetric profile was last seen in 1996. Since then
only symmetric profiles are present. Thus it is very unlikely that the
spectral state of the last 8 years correspond to a V/R phase.
We conclude that the density wave faded away before the dissipation of the
disc, perhaps because the disc became too tenuous to support a density wave.

Some spectra show the depression between the double peak profile extending
below the stellar continuum, reminiscent of the {\em shell} profile. These
types of lines are explained by partial absorption of the central star by
the circumstellar disc as a consequence of a high inclination angle 
\citep[see e.g.][]{humm00}. However, the profiles seen in \ls\ cannot be
considered as proper  shell lines because they only occurred during the
optical minimum. If ascribed to absorption by the disc itself then we
should expect the central depression to become more apparent as the extent
of the disc increases, i.e, as the  EW(H$\alpha$) increases. No such trend
is seen (Fig.~\ref{havarew}). In addition, the width of the central
reversal is considerably broader ($FWHM \approx 200-400$ km s$^{-1}$) than
the typical value of shell profiles ($FWHM \simless 50$ km s$^{-1}$).
Finally, none of the spectra of \ls\ fulfil the criterion given by
\citet{hanu96} that in order for a profile to have shell characteristics
the ratio $I_{\rm p}/I_{\rm cd}$ should be larger than 1.5.  Therefore the
apparent shell profiles in \ls\ are likely to be due to the 
photospheric absorption line, which combines with a weak
double-peaked emission.

Significant changes are apparent in the distance between the peaks
and the strength of the line.  The peak separation correlates with the
intensity of the H$\alpha$ line (Fig.~\ref{havarew}). As the EW(H$\alpha$)
increases, the distance between peaks decreases.
Interpreting the peak separation ($\Delta_{\rm peak}$) as the outer radius
($R_{ \rm out}$) of the emission line forming region \citep{huan72}

\begin{equation}
\frac{R_{ \rm out}}{R_*}=\sqrt{\frac{2 v \sin i}{\Delta_{\rm peak}}} 
\label{huang}
\end{equation}

\noindent we conclude that lower velocities of the emitting components
occur when the disc has developed, i.e, when the EW(H$\alpha$) is large.
Despite a moderate increase of the EW(H$\alpha$) the absence of
asymmetries indicates that fast radial displacements do not take place
during the first instances of the formation of the disc.

The \ion{He}{i}~6678\AA\ line also shows V/R variability. In  general, it
imitates the behaviour of the H$\alpha$ line. Since metallic lines are
generated at smaller disc radii than the hydrogen lines \citep{humm95,
jasc04}, the asymmetry of the \ion{He}{i} line profiles indicates that the
internal changes of the disc are global, affecting its entire structure.

\subsection{Variability time scales}

Although the spectroscopic data are distributed irregularly over the period of
the reported observations, the smooth
variations of the H$\alpha$ equivalent width and infrared brightness
indicate that structural changes in the circumstellar disc of \ls\ occur
on time scales of years. 

Table~\ref{loss} shows the typical time scales associated with disc
variability for a number of Be/X-ray binaries: $T_{\rm disc}$ is the
typical duration of formation/dissipation of the circumstellar disc and
$T_{\rm V/R}$ represents the quasi period for $V/R$ variability. $T_{\rm
disc}$ exhibits a good correlation with the orbital period. Systems with
narrow orbits tend to show faster disc growth and dissipation cycles,
while slower evolutionary time scales are associated with long orbital
periods. 
This is in agreement with the disc truncation model \citep{okaz01},
which suggests a direct relationship between
the size of the disc and the orbital period  \citep[see also][for
observational evidence in this respect]{reig97b}.  Within the framework of
the global one-armed oscillation model, the viscous excitation of a
density wave is associated with longer time scales when the disc is larger
\citep{okaz00}.

 Although the orbital period of \rosat/\ls\ is unknown, its
classification as a persistent system \citep{reig99} with a 
the relatively long
spin period implies that it must be long. A $P_{\rm orb}=150-200$ d is
estimated from the $P_{\rm spin}-P_{\rm orb}$ diagram \citep{corb86}. For
\object{LS I +61 235} Corbet's diagram implies $P_{\rm orb}\simmore 200$
d.  The typical time scales associated with the evolution of the
circumstellar disc in \ls\ are rather long  --- the EW(H$\alpha$) had not
recovered from 
the initial pre-disc-loss phase values nine years later --- and
agree with those of Be/X-ray binaries with wide orbits.
In contrast, changes originated by the density wave are much faster.
In \ls, the time elapsed between the slightly blue-dominated
line of the first spectrum and the strongly red-dominated line of the
second spectrum of Fig.\ref{havar} is just 3 months.


None of the three Be/X-ray binaries that have gone through disc-loss
phases and for which there is a good optical follow-up coverage, namely,
\object{X Per} in 1990 \citep{clar01}, \object{4U 0115+63} in 1997
\citep{negu01} and \object{1A 0535+262} in 1998 \citep{haig04}, exhibited
asymmetric profiles during the initial stage of disc growth. After the
disc loss phase  the first asymmetric profile did not occur until the
EW($H\alpha$) was $\sim 6-7$ \AA\ in \object{4U 0115+63} ($P_{\rm
orb}=24.3$ d), $\sim 7-10$ \AA\ in \object{1A 0535+262} ($P_{\rm orb}=111$
d) and  $\sim 10-12$ \AA\ in \object{X Per} ($P_{\rm orb}=250$ d). In our
monitoring of \ls/\rosat, asymmetric profiles are associated with  the
largest values of the EW(H$\alpha$). Below 8 \AA, only symmetric profiles
are observed. The peak separation of the latest spectra imply a disc
radius of $\sim$2 $R_*$, assuming a Keplerian disc and Eq.(1).

In summary, the correlation of $T_{\rm disc}$ and the orbital period
provides further observational evidence for the interaction of the neutron
star with the circumstellar disc of its Be star's companion, whilst the
relationship betwen the EW(H$\alpha$) at which the first asymmetry appears
with the orbital period implies that the density oscillations do not
become observable until the disc has reached a critical size or density.

\section{Conclusion}

We have monitored the Be/X-ray binary \ls\ for the last 10 years. The
observations coincided with a period of low optical/IR activity,
characterised by the likely loss of the Be star's circumstellar disc and
subsequent reformation. Since 2001 the envelope has been 
gradually growing as
indicated by the increase of the equivalent width and the narrowing of the
peak separation of the split H$\alpha$ line. The time scales for
structural changes in the circumstellar disc of \rosat/\ls\ compares
favourably with
those of Be/X-ray binaries with long orbital periods.  While the
formation/dissipation of the disc may last for several years, the line
profile changes are much faster and, in general, depends on the duration
of the active phase. The disappearance of the V/R varibility before
the dissipation of the disc and the lack of asymmetric profiles of the
latest observations even though the equivalent width of the H$\alpha$ line
has increased up to $\sim 6$ \AA\ confirms the fact that the effects of
the density perturbation do not manifest themselves until the disc  is
fully developed. By studying the characteristic variability time scales
of a number of Be/X-ray binaries we have found further observational
evidence of the influence of the neutron star on the envelope of the Be
star.

\begin{acknowledgements}

The authors thank M. Brotherton and P. Nandra for providing the Kitt Peak
spectrum. The Kitt Peak National Observatory, a division of the National
Optical Astronomy Observatories, is operated by the Association of
Universities for Research in Astronomy, Inc. under cooperative agreement
with the National Science Foundation. IN is a researcher of the programme
{\em Ram\'on y Cajal}, funded by the Spanish Ministerio de Educaci\'on y
Ciencia and the University of Alicante, with partial support from the
Generalitat Valenciana and the European Regional Development Fund
(ERDF/FEDER). This research is partially supported by the MEC through
grant ESP-2002-04124-C03-03.  Skinakas Observatory is a collaborative
project of the University of Crete, the Foundation for Research and
Technology-Hellas and the Max-Planck-Institut f\"ur Extraterrestrische
Physik. The WHT spectrum was obtained as part of the ING service
programme. Based in part on observations made at Observatoire de Haute
Provence (CNRS), France. The Carlos Sanchez Telescope is operated at the
Teide Observatory by the Instituto de Astrof\'{\i}sica de Canarias.

\end{acknowledgements}

\end{document}